\documentclass[aoas]{imsart}

\RequirePackage{amsthm,amsmath,amsfonts,amssymb}
\RequirePackage[authoryear]{natbib}

\startlocaldefs
\usepackage{indentfirst}
\usepackage{appendix}
\usepackage{epstopdf}
\usepackage{amsbsy}
\usepackage{amsmath}
\usepackage{amssymb}
\usepackage{epsfig}
\usepackage{multicol}
\usepackage{multirow}
\usepackage{verbatim}
\usepackage{graphicx}
\usepackage{float}
\usepackage{booktabs}
\usepackage{caption}
\usepackage{subfigure}
\usepackage{threeparttable}
\usepackage{algorithm}
\usepackage{algorithmic}
\usepackage{array}
\usepackage{natbib}
\DeclareMathOperator*{\argmin}{argmin}
\DeclareMathOperator*{\argmax}{argmax}
\usepackage[colorlinks,linkcolor=blue,anchorcolor=blue,citecolor=blue]{hyperref}
\usepackage{graphicx}
\usepackage{amssymb, amsmath, bm, natbib, mathrsfs}

\newtheorem{theorem}{Theorem}
\newtheorem{lemma}{Lemma}

\endlocaldefs

\begin{document}

\begin{frontmatter}
\title{On Estimation of Optimal Dynamic Treatment Regimes with Multiple Treatments for Survival Data-With Application to Colorectal Cancer Study}
\runtitle{DTR with Multiple Treatments for Survival Data}

\begin{aug}

\author[A]{\fnms{Zhishuai}~\snm{Liu}},
\author[B]{\fnms{Zishu}~\snm{Zhan}}\footnote{Zhishuai Liu and Zishu Zhan contributed equally to this work.},
\author[E]{\fnms{Jian}~\snm{Liu}},
\author[C]{\fnms{Danhui}~\snm{Yi}},
\author[D]{\fnms{Cunjie}~\snm{Lin}\ead[label=e4]{lincunjie@ruc.edu.cn}}
\and
\author[E]{\fnms{Yufei}~\snm{Yang}\ead[label=e5]{yyf93@vip.sina.com
}}
\address[A]{Department of Biostatistics \& Bioinformatics, Duke University}

\address[B]{Department of Biostatistics, School of Public Health, Southern Medical University}

\address[C]{School of Statistics, Renmin University of China}

\address[D]{Center for Applied Statistics and School of Statistics, Renmin University of China\printead[presep={,\ }]{e4}}

\address[E]{Xiyuan Clinical Medical College, Beijing University of Chinese Medicine\printead[presep={,\ }]{e5}}
\end{aug}

\begin{abstract}
Dynamic treatment regimes (DTR) are sequential decision rules corresponding to several stages of intervention. Each rule maps patients' covariates to optional treatments. The optimal dynamic treatment regime is the one that maximizes the mean outcome of interest if followed by the overall population. Motivated by a clinical study on advanced colorectal cancer with traditional Chinese medicine, we propose a censored C-learning (CC-learning) method to estimate the dynamic treatment regime with multiple treatments using survival data. To address the challenges of multiple stages with right censoring, we modify the backward recursion algorithm in order to adapt to the flexible number and timing of treatments. For handling the problem of multiple treatments, we propose a framework from the classification perspective by transferring the problem of optimization with multiple treatment comparisons into an example-dependent cost-sensitive classification problem. With classification and regression tree (CART) as the classifier, the CC-learning method can produce an estimated optimal DTR with good interpretability. We theoretically prove the optimality of our method and numerically evaluate its finite sample performances through simulation. With the proposed method, we identify the interpretable tree treatment regimes at each stage for the advanced colorectal cancer treatment data from Xiyuan Hospital.
\end{abstract}

\begin{keyword}
\kwd{Dynamic treatment regime}
\kwd{multiple treatments} 
\kwd{classification perspective}
\kwd{survival data}
\end{keyword}
\end{frontmatter}

\section{Introduction}
\label{introduction}
The idea of personalized medicine is increasingly of interest today, which points out that clinicians should tailor the most suitable strategies on an individual level using patient-specific data sources to optimize some outcomes. As a key branch of personalized medicine, dynamic treatment regimes (DTR), formalize a series of sequential decision rules across multiple stages of interference, in which each rule takes the patients' historical treatments and covariates up to that stage as input and outputs the recommended treatment in the next stage. The ultimate goal is to establish an optimal dynamic treatment strategy that if followed by the overall population can achieve the maximum benefit on average.

There is a substantial statistical literature on methods to identify an optimal DTR \citep{Zhang:2013,Zhao:2015,Zhang:2018,Luckett:2020,chen:2023}. However, estimating optimal DTRs can be challenging when there are more than two treatments at each stage of the intervention due to the complexity of multiple treatment comparisons, and further methodological developments are required. For example, \cite{Tao:2017} proposed an adaptive contrast weighted learning (ACWL) method to estimate the optimal dynamic treatment regime with multiple treatments from the classification perspective based on the augmented inverse probability weighted (AIPW) estimator. However, ACWL recasts the problem of optimization with multiple treatment comparisons as a weighted classification problem, which is example-dependent cost-sensitive \citep{Lin:2019} and may result in unsatisfactory performance without adequate samples. 
Following ACWL, \cite{Tao:2018} proposed a tree-based reinforcement learning (T-RL) method. It takes the advantages of the double robustness of the AIPW estimator and the interpretability of tree-type regimes. Recently, \cite{Ma:2023} proposed a group outcome weighted learning (GROWL) to estimate the latent structure in the treatment space and the optimal group-structured ITRs through a single optimization, which is worth further extending to learn group structures for multi-stage dynamic treatment regimes. But these methods are not suitable for survival data.

For chronic diseases such as cancer, survival data often arises, and developing an optimal DTR is important under survival data framework. There are two main challenges in estimating the optimal dynamic treatment regime when the outcome is the survival time. First, the survival time is subject to right censoring and the outcome of interest may not be observed for some patients by the end of the study. Second, 
the total number of treatment stages may vary among patients due to failure or dropping out of the study. We note that most existing approaches for estimating dynamic treatment regimes are based on backward recursion, which is a dynamic programming technique commonly used in reinforcement learning \citep{Bellman:1966}. This requires one to start finding the optimal regime at the last stage, but the optimal treatment at the last stage is not well defined for survival outcome and therefore the backward recursion is not applicable. To solve these challenges, \cite{Goldberg:2012} presented a censored Q-learning algorithm that is adjusted for censored data using inverse probability of censoring weighting, and developed a methodology for solving backward recursion by introducing an auxiliary multistage decision problem when the number and timing of stages are flexible. But the method requires positing parametric models for the survival time and lacks robustness to model misspecification. 
 \cite{Simoneau:2020} extended the dynamic weighted ordinary least squares (dWOLS) approach \citep{Wallace:2015} to the censored data. This method, named dynamic weighted survival modeling (DWSurv), is doubly robust and easily implementable. However, both of the methods rely heavily on the correct specification of the regression model because of the nature of Q-learning and A-learning. \cite{Zhang:2022} further extended DWSurv to estimate optimal DTRs of the censored data with multiple treatment options, which also inherits the shortcomings of DWSurv. In addition, \cite{Xue:2021} proposed a multi-category angle-based learning framework that takes both the survival outcome and multiple treatments into account. This method geometrically explains treatment differences, but lacks practical interpretability compared to tree-based decisions. Additionally, the use of angle classifiers is inherently limited to a specific class of dynamic treatment strategies, and the optimal strategy can only be sought within this constrained class. If the true optimal dynamic treatment strategy does not conform to an angular form, this method loses both interpretability and accuracy. Recently, 
 \cite{Cho:2023} developed a general dynamic treatment regime estimator for censored data, which allows the failure time to be conditionally independent of censoring and dependent on the treatment decision times. However, the estimator is constructed using generalized random survival forests, which lacks of interpretability.

In this paper, we propose a censored C-learning (CC-learning) method to extend the C-learning framework \cite{Zhang:2018} to multiple treatments and survival data to estimate the optimal dynamic treatment regime maximizing the mean survival time. The proposed method transfers the optimization problem into an example-dependent cost-sensitive classification problem \citep{Elkan:2001}. Then existing powerful classification techniques combined with the data space expansion technique \citep{Abe:2004} can be used to solve the classification problem and estimate the optimal dynamic treatment regime. Also, our proposed method allows the optimization step of the decision rule to be decoupled from the modeling process through a two-step approach?. Furthermore, CC-learning not only address the challenges posed by survival data and multiple classes of treatment regimens for the estimation of dynamic treatment strategies, but also take into account interpretability. 

The content of the paper is summarized as follows: 
Section \ref{data} introduces the background of the motivation data.
Section \ref{Methodology} introduces the mathematical framework and the estimation algorithm. Section \ref{Simulation} demonstrates the finite sample performance through numerical studies. We apply the method to the advanced colorectal cancer data in section \ref{Data analysis}. We conclude the study in Section \ref{Conclusions}. All the proofs are provided in the Appendix.

\section{Data}
\label{data}
This research is motivated by a clinical observational study on advanced colorectal cancer conducted by the China Academy of Chinese Medical Sciences, Xiyuan Hospital. Colorectal cancer is one of the most common malignant tumors in the world, and its morbidity and mortality have remained high for many years \citep{Wild:2020}. The routine treatment of advanced colorectal cancer is chemotherapy combined with molecular targeted therapy and biological immunotherapy, which is pure Western medicine. Although this treatment effectively prolongs the survival of patients, the adverse reactions it causes, as well as the symptoms and complications of advanced colorectal cancer itself, often make patients suffer intolerably. Furthermore, conventional Western medicine treatments do not cover all populations, and for patients who are physically weak or economically disadvantaged, more appropriate treatment options are urgently needed.

Traditional Chinese medicine has been following the treatment principles of patient-centered care and individualized diagnosis and treatment, which aligns with precision medicine naturally. According to studies of \cite{Wang:2020}, \cite{Yeh:2020} and \cite{Shao:2019}, Chinese medicine plays a vital role in prolonging survival time and improving living quality in the treatment of advanced colorectal cancer. Of course, 
not all patients can benefit from it due to heterogeneity in treatment effect. Hence, in addition to Western medicine and Chinese medicine, an integrated treatment approach that combines both Chinese medicine and Western medicine is also considered for cancer treatment. Chinese medicine has long been an active part of cancer care, across various disease types and either alone or in combination with conventional cancer therapies in China \citep{Li:2017}. Chinese medicine treatment can play a dominant role, cooperate with chemotherapy, or be integrated with Western medicine treatment at different stages of the disease for advanced colorectal cancer patients. Therefore, the treatment of advanced colorectal cancer cannot rely solely on either Chinese or Western medicine, but should be based on the different stages of disease development and the characteristics of patients' diseases and demographics.

\begin{figure}[htbp]
\centering
\includegraphics[width=12cm,height=11cm]{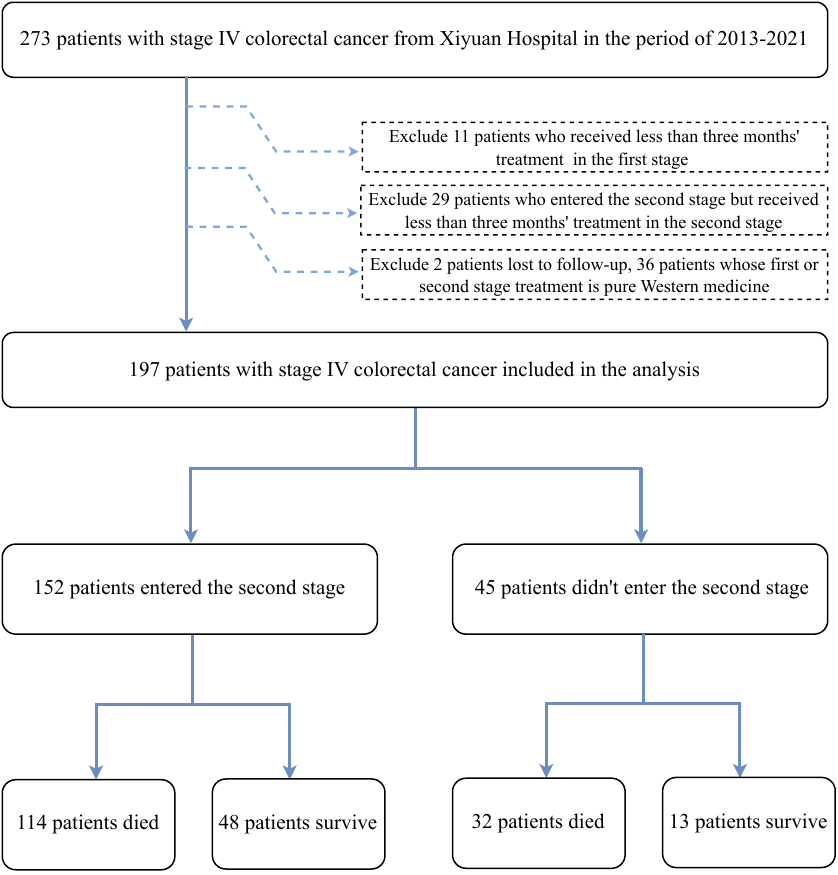}
\caption{The flowchart of data preparation.}
\label{flowchart}
\end{figure}

In this study, data collection started on January 1, 2013 and ended on January 1, 2021.
In the dataset, 273 patients with stage IV colorectal cancer received two-stage treatments. The data preparation flowchart is presented in Figure \ref{flowchart}. After data processing, 197 patients are left, among which 152 patients entered the second stage and 45 did not due to death or loss of follow-up, which means that the number of treatment stages varies across patients. Among the 197 patients, 51 are censored, thus the censoring rate is 25.9\%. At each stage, there are three treatment options, which are pure traditional Chinese medicine ($A$=0), combined treatment of traditional Chinese and Western medicine ($A$=1) and Western medicine assisted with traditional Chinese medicine ($A$=2), respectively.  
There is one continuous variable ${\rm Time}_1$ (treatment time at the first stage), and eight categorical covariates: Gender, Age, Site, Metastasis, Genotyping, ECOG1, ECOG2 and Stage. More details can be found in Section \ref{Data analysis}. The researchers seek interpretable strategies based on these covariates. Thus, the colorectal cancer data involves the three aspects we mentioned before:
First, the patients receive multi-stage therapies and there are multiple treatment options at each stage; Second, some patients are censored during the follow-up, which yields flexible number and timing  of treatment stages; Third, an optimal dynamic treatment regimes with good interpretability are required, especially for clinicians to improve the treatment and prognosis of the disease.

Therefore, this study aims to find the optimal dynamic treatment regime maximizing mean survival time based on the advanced colorectal cancer patients' physiological, disease, biological and demographic characteristics. Considering a two-stage dynamic treatment strategy and three available treatments, our proposed method is used to obtain the interpretable tree treatment strategies at each stage and select the decision variables. This provides empirical references for the subsequent treatment of advanced colorectal cancer, thereby benefiting the patients.

\section{Methodology}
\label{Methodology}

\subsection{Set-up and notations}

\begin{figure}
    \centering
    \includegraphics{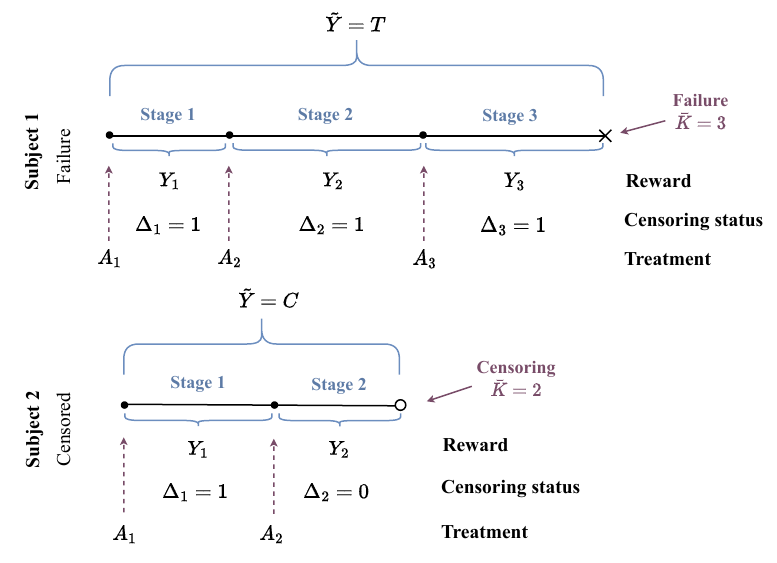}
    \caption{An example of two types of subjects}
    \label{fig:eg}
\end{figure}

We consider a multi-stage decision making process with $K$ ($K\geq 2$) stages of intervention. The covariates collected at the beginning of stage $k$ is denoted as $X_k\in\mathcal{X}_k\subset{\mathbb{R}^p}$ and we define the covariates history up to stage $k$ as $\bar{X}_k=(X_1,\cdots,X_k)\in \bar{\mathcal{X}}_k=\mathcal{X}_1\times\cdots\times \mathcal{X}_k$. Correspondingly, the observed covariates and covariates history are denoted as $x_k$ and $\bar{x}_k=(x_1,\cdots, x_k)$. The treatment received at the beginning of stage $k$ is denoted as $A_k, A_k \in \mathcal{A}_{k}$, where $\mathcal{A}_{k}$ is the set of treatment options at stage $k$ and we assume that there are at least three treatment options at each stage. The history of treatment up to stage $k$ is $\bar{A}_k=(A_1,\cdots,A_k)\in \bar{\mathcal{A}}_k=\mathcal{A}_1\times\cdots\times \mathcal{A}_k$. The observed values of $A_k$ and $\bar{A}_k$ are denoted as $a_k$ and $\bar{a}_k=(a_1,\cdots, a_k)$, respectively. The reward of stage $k$ is $Y_k$, which represents the time span between the $k$-th treatment and the $(k+1)$-th treatment. The ultimate outcome of interest $T \in \mathbb{R}^+$ is the time from the beginning of the first treatment to the event of failure, which is subject to right censoring. Denote the censoring time as $C\in\mathbb{R}^+$ and the observed ultimate outcome as $\tilde{Y}=\min\{T,C\}$. Following \cite{Goldberg:2012}, we assume that the censoring is independent of both covariates and failure time for simplicity. Let $\Delta_k=I(\sum_{j=1}^kY_k \leq C)$ denote the censoring indicator at stage $k$. If the censoring happens at stage $k$, then $\Delta_{k-1}=1$ and $\Delta_{k}=0$. It is worth noting that failure event and censoring can occur at any stage. Once either of them happens at stage $k$, the treatments and covariates after stage $k$ are not available. We denote the number of stages of treatment received by a patient as $\bar{K}$, which ranges from 1 to $K$. If the patient experiences the failure event right after the first-stage treatment, we observe that $\bar{K}=1$. If the event occurs after completing all $K$ stages of treatment, the observed number of stages of treatment is $\bar{K}=K$. Now we redefine the reward of last stage for the patient with $\bar{K}$ stages as $Y_{\bar{K}}=\min\{T-\sum_{k=1}^{\bar{K}-1}Y_k, C-\sum_{k=1}^{\bar{K}-1}Y_k\}$. Then the individual trajectories are 
\begin{eqnarray*}
\left\{X_{1}, A_{1}, \Delta_{1}, Y_{1}, \cdots, X_{\bar{K}}, A_{\bar{K}}, \Delta_{\bar{K}}, Y_{\bar{K}}, \tilde{Y} \right\}, 
\end{eqnarray*}
where $\tilde{Y}=\Delta_{\bar{K}}T+(1-\Delta_{\bar{K}})C=\sum_{k=1}^{\bar{K}}Y_{k}$. Figure \ref{fig:eg} displays two types of subjects with different number of stages and censoring statuses.

Suppose that we have $n$ independent patients with index $i=1,\cdots, n$ and the observed data are
\begin{eqnarray*}
\left\{X_{1i}, A_{1i}, \Delta_{1i}, Y_{1i}, \cdots, X_{\bar{K}i}, A_{\bar{K}i}, \Delta_{\bar{K}i}, Y_{\bar{K}i}, \tilde{Y}_i \right\}_{i=1}^n, 
\end{eqnarray*}
$\tilde{Y}_i=\Delta_{\bar{K}i}T_i+(1-\Delta_{\bar{K}i})C_i=\sum_{k=1}^{\bar{K}i}Y_{ki}$.
Sometimes, only the overall survival time $\tilde{Y}$ is observed while
the intermediate rewards $Y_k$ are not recorded, in this case, the observed data are $\left\{X_{1i}, A_{1i}, \cdots, X_{\bar{K}i}, A_{\bar{K}i}, \Delta_{i}, \tilde{Y}_i \right\}_{i=1}^n,$ where $\Delta_{i}=I(T_i \leq C_i)$.

\subsection{Backward induction and trajectory completing}
Denote $g_k:\bar{\mathcal{X}}_k\times \bar{\mathcal{A}}_{k-1}\rightarrow \mathcal{A}_k$ as the decision rule in stage $k$ that maps the subject's historical information $L_k=(\bar{X}_k,\bar{A}_{k-1})$ to the treatment assignment space, and a dynamic treatment regime $g = \left(g_1,\cdots, g_K\right)$ is an ordered set of decision rules. We consider a potential outcomes framework and let $\tilde{Y}^{*}(g)$ be the potential outcome corresponding to a regime $g$. Then the optimal dynamic treatment regime $g^{opt} = (g^{opt}_{1},\cdots, g^{opt}_{K})$ is defined as the one that if followed by the overall population can achieve the longest mean survival time. In other words, $g^{opt}$ satisfies
\begin{equation}
\label{definition}
E\{\tilde{Y}^*(g^{opt})\} \geq E\{\tilde{Y}^*(g)\}, ~~\forall~ g ~\in \mathcal{G},
\end{equation}
here $\mathcal{G}=\mathcal{G}_1\times\cdots\times\mathcal{G}_K$ and $\mathcal{G}_k$ is a pre-specified class of functions. Similar to \cite{Schulte:2014}, we make the following assumptions to make  $g^{opt}$ identifiable using the observed data: (i) the consistency assumption, (ii) the stable unit treatment value assumption (SUTVA), (iii) the no unmeasured confounders assumption (NUCA) and (iv) positivity assumption. Then the optimal dynamic treatment regime can be expressed by the observed data through backward induction. Specifically, when the number of stages is the same for all observations, we define the $Q$-function at the $K$-th stage as
\begin{eqnarray*}
\label{Q_K}
Q_K(l_K,a_K)=E(\tilde{Y}|{L}_K=l_K, A_K=a_K),
\end{eqnarray*}
where $l_K = (\bar{x}_K, \bar{a}_{K-1})$. Then the optimal treatment regime at stage $K$ is
\begin{eqnarray*}
\label{g_K1}
g_K^{opt}=\argmax_{a_K \in \mathcal{A}_K}Q_K({l}_{K},a_K).
\end{eqnarray*}
Correspondingly, the value function at stage $K$ is $V_K(l_K)=\max_{a_K \in \mathcal{A}_K}Q_K({l}_K,a_K).$
Recursively, we define the $Q$-function at stage $k=K-1,\cdots,1$ as
\begin{eqnarray*}
\label{Q_k}
Q_k({l}_k, a_k)=E\left\{V_{k+1}(L_{k+1})|{L}_k=l_k, A_k=a_k\right\}, 
\end{eqnarray*}
where $l_k=(\bar{x}_k, \bar{a}_{k-1})$,
and the corresponding value function 
\begin{eqnarray*}
\label{Value function}
V_k({l}_k)=\max_{a_k \in \mathcal{A}_k}Q_k({l}_k,a_k).
\end{eqnarray*}
Then the optimal treatment regime at the stage $k$ satisfies
\begin{eqnarray*}
\label{g_k1}
g_k^{opt}=\argmax_{a_k \in \mathcal{A}_k}Q_k({l}_k,a_k).
\end{eqnarray*}
We refer to \cite{Schulte:2014} for more details on the backward induction framework. In the previous, we conduct the backward induction directly based on the definition of the $Q$-function and the value function, we denote it as the D-method with ``D'' representing ``Direct''. We can also provides a way of backward induction based on regret function defined in the following theorem, we denote id as the R-method with ``R'' representing ``Regret''.

\begin{theorem}
\label{th1}
Under the assumptions (i), (ii), (iii) and (iv) mentioned above, the value functions can be derived through the following equation: 
\begin{equation}
E\left\{V_{k+1}\left(L_{k+1}\right)+\left(Q_{k}\left(L_{k}, g_{k}^{opt}\right)-Q_{k}\left(L_{k}, A_{k}\right)\right) \bigg| L_{k}\right\}=V_{k}\left(L_{k}\right),    
\end{equation}
where $k=K, \cdots, 1$ and $V_{K+1} \equiv \tilde{Y}$.
\end{theorem}
Theorem \ref{th1} indicates that the R-method is theoretically equivalent to the D-method, they are significantly different in practice. It is discussed in \cite{Huang:2015} and \cite{Tao:2017} that the R-method can reduce the accumulated bias during the backward induction and is more robust against the $Q$-function misspecification. But the robustness is limited, especially when the $Q$-function is severely misspecified. Furthermore, the loss $Q_{k}(L_k, g_k^{opt}) - Q_{k}(L_k, A_k)$ can be negative, which severely violates the definition. The D-method can avoid the problem of negative loss, though it is more sensitive to the misspecification of the $Q$-function to some extent. 

However, as mentioned in section \ref{introduction}, the traditional backward induction method mentioned above cannot be applied directly to survival data due to the flexible number of stages: (i) when the number of stages is varied, it is not clear how to perform the base step of the recursion, (ii) some of the trajectories may be incomplete due to the censoring and the reward is not known for the stage at which censoring occurs. Similar to \cite{Goldberg:2012}, we construct an auxiliary problem for allowing the implementation of backward induction with flexible numbers and timing of stages. Specifically, when the failure event or censoring occurs in stage $k < K$, we define $X'_j= X_j, A'_j= A_j, Y'_j= Y_j$ and $\Delta'_j=\Delta_j$ for $j\leq k$; let $X'_j=\emptyset, Y'_j=0, \Delta'_j=\Delta_k, \tilde{Y}' = \tilde{Y}$ and generate $A'_j$ uniformly from $\mathcal{A}_j$ for $k+1\leq j\leq K$. Then the modified data trajectory is $\left\{X'_1, A'_1, \Delta'_1, Y'_1,\cdots, X'_K, A'_K, \Delta'_K, Y'_K,\tilde{Y}'\right\}$, or $\left\{X'_1, A'_1,\cdots, X'_K, A'_K, \Delta',\tilde{Y}'\right\}$ if the intermediate rewards were not recorded, where $\tilde{Y}'=\sum_{k=1}^KY_k'$, $\Delta'=\Delta_k$. We refer to estimating the optimal dynamic treatment regime using the modified data as the auxiliary problem and using the unmodified data as the original problem. Then it can be proved the following Lemma 

\begin{lemma}\label{lem1}
    (The auxiliary problem) Under the assumption (iv), 
    \begin{equation}
\label{equivalence}
E_g\left(\sum_{j=1}^{K} Y_{j}^{\prime} \mid X_{1}=x_{1}\right)=E_{0,g}\left\{\sum_{j=1}^{\bar{K}} Y_{j} \mid X_{1}=x_{1}\right\},
\end{equation}
where $E_g$ and $E_{0,g}$ are expectations with respect to the distributions of the modified trajectory and the original observed trajectory following the strategy $g$, respectively.
\end{lemma}

Lemma \ref{lem1} relates the value function in the auxiliary problem to the original problem, which means that the optimal dynamic treatment regime of the original problem can be obtained by optimizing the auxiliary problem.  Hence in the following, we construct the optimal dynamic treatment regime estimation framework for the auxiliary problem. Without any risk of ambiguity, we omit the subscript of expectation ``$E$'' in the rest of the paper. The proof of Lemma \ref{lem1} follows directly from Lemma 4.1 in \cite{Goldberg:2012}.

\subsection{Censored C-learning}
\label{Clearning}
In this section, we extend the C-learning proposed by \cite{Zhang:2018} to accommodate the censored cases with the auxiliary problem. We need further notations to construct the recursive optimization framework from the classification perspective. Suppose that there are $m_k$ treatments in stage $k$, that is $\mathcal{A}_k=\{1,\cdots,m_k\}$. For $1\leq k\leq K$, denote $Q_k(L_k,a)$ as $Q^{k}_{a}(L_k)$, $1 \leq k \leq K$, then the $Q$-functions corresponding to $m_k$ treatments are $Q^{k}_{1}(L_k), \cdots,Q^{k}_{m_k}(L_k)$, respectively. Denote the $s$-th order statistics of the $Q$-function at stage $k$ as $Q^{k}_{(s)}(L_k)$, then we have $Q^{k}_{(1)}(L_k) \leq \cdots \leq Q^{k}_{(m_k)}(L_k)$. Let $l^{k}_{s}(L_k), 1 \leq s \leq m_k$ be the corresponding treatment of $Q^{k}_{(s)}(L_k)$, and define the contrast function as $C^{k}_{s}(L_k) = Q^{k}_{(m_k)}(L_k)-Q^{k}_{s}(L_k), 1 \leq s \leq m_k$, which represents the difference in effect between treatment $s$ and the optimal treatment $l^{k}_{m_k}(L_k)$. 
The recursive optimization framework of the optimal dynamic treatment regime is given by Theorem \ref{th2}. The proofs are given in the Appendix.

\begin{theorem}
\label{th2}
Under the assumptions (i), (ii), (iii) and (iv) mentioned above, the optimal dynamic treatment regime $g^{opt}=(g^{opt}_{1}, \cdots, g^{opt}_{K})$ satisfies
\begin{equation}
g^{opt}_{k}=\underset{g_{k} \in \mathcal{G}_{k}}{\arg \min } E\left\{C_{g_{k}}^{k}\left(L_{k}\right) I(g_{k}\ \neq l_{m_{k}}^{k}\left(L_{k})\right)\right\},~~~k=1,\cdots, K.
\end{equation}
\end{theorem}

According to Theorem \ref{th2}, estimating the optimal dynamic treatment regime is equivalent to solving a weighted classification problem. Specifically, we can refer to the contrast functions $C^k_{g_{k}}\left(L_{k}\right)$ as the misclassification cost, and $l_{m_{k}}^{k}\left(L_{k}\right)$ as the labels. Hence the optimal dynamic treatment regime can be perceived as the one that minimizes the mean misclassification cost.  


The results of Theorem \ref{th1} and Theorem \ref{th2} provide us with the recursive optimization framework for estimating the optimal dynamic treatment regime with multiple treatments for survival data. Accordingly, we start with the last stage, and then estimate the optimal regime at each stage in reverse order. Specifically, the optimal treatment regime at stage $k$ is estimated by
\begin{eqnarray}
\label{classification}
\hat{g}^{opt}_k=\argmin_{g_k\in \mathcal{G}_k}\sum_{i=1}^n\left \{\hat{C}_{g_k}^k(L_{ki}) I(g_k \neq \hat{l}_{m_k}^k(L_{ki})) \right \},~~~~ k=K,\cdots, 1,
\end{eqnarray}
where  $\hat{C}_{g_k}^k(L_{ki})$ and $\hat{l}_{m_k}^k(L_{ki})$ are estimators of $C_{g_k}^k(L_{ki})$ and $l_{m_k}^k(L_{ki})$, respectively. Here, we use the double robust AIPW estimator for the contrast function.

First, for data trajectory $\left\{X_1, A_1, \Delta_1, Y_1, \cdots, X_K, A_K, \Delta_K, Y_K, \tilde{Y}\right\}$, denote $\mathcal{P}_n$ as the sample mean, we have
\begin{align}
\label{C-AIPW1}
\hat{C}_{s}^{k}(L_k)=&\mathcal{P}_n\Bigg\{\frac{I(A_k=\hat{l}_{m_k}^k(L_k))\Delta_k}{\hat{\pi}_{\hat{l}_{m_k}^k(L_k)}^k(L_k)\hat{S}_C(\sum_{j=1}^kY_k)}\hat{V}_{k+1}(L_{k+1})+\Bigg[1-\frac{I(A_k=\hat{l}_{m_k}^k(L_k))}{\hat{\pi}_{\hat{l}_{m_k}^k(L_k)}^k(L_k)}\Bigg]\hat{Q}^{k}_{\hat{l}_{m_k}^k(L_k)}(L_k) \\
& -\notag\left [\frac{I(A_k=s)\Delta_k}{\hat{\pi}_{s}^k(L_k)\hat{S}_C(\sum_{j=1}^kY_k)}\hat{V}_{k+1}(L_{k+1}) + \left\{ 1-\frac{I(A_k=s)}{\hat{\pi}_{s}^k(L_k)}\right\}\hat{Q}^k_s(L_k)\right ]\Bigg\},
\end{align}
\begin{eqnarray*}
\label{l-AIPW}
\hat{l}_{m_k}^{k}(L_k)=\argmax_{s \in \mathcal{A}_k}\mathcal{P}_n\left[\frac{I(A_k=s)\Delta_k}{\hat{\pi}_{s}^k(L_k)\hat{S}_C(\sum_{j=1}^kY_k)}\hat{V}_{k+1}(L_{k+1})+\left \{1-\frac{I(A_k=s)}{\hat{\pi}_{s}^k(L_k)}\right\}\hat{Q}^{k}_{s}(L_k)\right],
\end{eqnarray*}
 where $\hat{V}_{k+1}(L_{k+1})$, $\hat{Q}^k_{a}(L_k)$, and $\hat{\pi}_{a}^k(L_k)$ are estimators of the value function $V_{k+1}(L_{k+1})$, the $Q$-function $Q^k_{A_k}(L_k)$, and the propensity score $\pi_{a}^k(L_k)=P(A_k=a|L_k)$, here, $a=s$ or $\hat{l}_{m_k}^k(L_k)$; $\hat{S}_C(\sum_{j=1}^kY_k)$ is the estimator for the survival function of the censoring time $C$. Specifically, we consider $\{\mathcal{Q}_1,\cdots, \mathcal{Q}_K \}$ as the space of the $Q$-functions. As mentioned above, when the failure event or censoring happens before stage $k$, we set $X_k=\emptyset$, $Q_k(L_k,A_k)=V_{K+1}=\tilde{Y}$. Otherwise, we estimate the $Q$-function at the stage $k$ as follows:
 \begin{eqnarray}
\label{Q-IPCW}
\hat{Q}_k(L_k,A_k)=\argmin_{\mathcal{Q}_k}\mathcal{P}_n\left[\left(V_{k+1}(L_{k+1})- Q_k(L_k,A_k)\right)^2\frac{\Delta_k}{\hat{S}_C(\sum_{j=1}^kY_k)}\right].
\end{eqnarray}
Then we can estimate the value function with the D-method 
\begin{eqnarray}
\label{M}
\hat{V}_{k}(L_k)=\max_{A_k\in \mathcal{A}_k}\hat{Q}_k(L_k, A_k), 
\end{eqnarray}
or the R-method
\begin{eqnarray}
\label{R}
\hat{V}_{k}(L_k)=\hat{V}_{k+1}(L_{k+1})+\hat{Q}_{k}(L_k, \hat{l}_{m_k}^k(L_k)) - \hat{Q}_{k}(L_k, A_k).
\end{eqnarray}
Besides, we adopt the softmax regression to estimate the propensity score $\pi_{a}^k(L_k)=P(A_k=a|L_k)$, that is 
\begin{eqnarray*}
\hat{\pi}_{A_k}^k(L_k)=\frac{\exp\left \{\hat{\gamma}_{A_k}^{k\top}\tilde{L}_{k0}+(\hat{\phi}_{A_k}^{k\top}\tilde{L}_{k1})A_k\right \}}{\sum_{a_k \in \mathcal{A}_k} \exp\left\{\hat{\gamma}_{a_k}^{k\top}\tilde{L}_{k0}+(\hat{\phi}_{a_k}^{k\top}\tilde{L}_{k1})a_k\right \}},
\end{eqnarray*}
where $\tilde{L}_{k0}$ and $\tilde{L}_{k1}$ are subsets of $L_k$, $\tilde{L}_{k0}$ are covariates in the main effect term and $\tilde{L}_{k1}$ are covariates in the interaction term. The survival function of the censoring time can be estimated by the Kaplan-Meier (KM) estimator.

Second, for data trajectory $\left\{X_1, A_1,\cdots, X_K, A_K, \Delta,\tilde{Y}\right\}$, we have
\begin{align}
\label{C-AIPW2}
\hat{C}_{s}^{k}(L_k)=&\mathcal{P}_n\Bigg\{\frac{I(A_k=\hat{l}_{m_k}^k(L_k))\Delta}{\hat{\pi}_{\hat{l}_{m_k}^k(L_k)}^k(L_k)\hat{S}_C(\tilde{Y})}\hat{V}_{k+1}(L_{k+1})+\Bigg[1-\frac{I(A_k=\hat{l}_{m_k}^k(L_k))}{\hat{\pi}_{\hat{l}_{m_k}^k(L_k)}^k(L_k)}\Bigg]\hat{Q}^{k}_{\hat{l}_{m_k}^k(L_k)}(L_k) \\
& -\notag\left [\frac{I(A_k=s)\Delta}{\hat{\pi}_{s}^k(L_k)\hat{S}_C(\tilde{Y})}\hat{V}_{k+1}(L_{k+1}) + \left\{ 1-\frac{I(A_k=s)}{\hat{\pi}_{s}^k(L_k)}\right\}\hat{Q}^k_s(L_k)\right ]\Bigg\},
\end{align}
\begin{eqnarray*}
\label{l-AIPW}
\hat{l}_{m_k}^{k}(L_k)=\argmax_{s \in \mathcal{A}_k}\mathcal{P}_n\left[\frac{I(A_k=s)\Delta_k}{\hat{\pi}_{s}^k(L_k)\hat{S}_C(\tilde{Y})}\hat{V}_{k+1}(L_{k+1})+\left \{1-\frac{I(A_k=s)}{\hat{\pi}_{s}^k(L_k)}\right\}\hat{Q}^{k}_{s}(L_k)\right],
\end{eqnarray*}
\begin{eqnarray}
\label{Q-IPCW2}
\hat{Q}_k(L_k,A_k)=\argmin_{ \mathcal{Q}_k}\mathcal{P}_n\left[\left(V_{k+1}(L_{k+1})- Q_k(L_k,A_k)\right)^2\frac{\Delta}{\hat{S}_C(\tilde{Y})}\right].
\end{eqnarray}
The estimations of the value function and the propensity score are the same as those of data trajectory $\left\{X_1, A_1, \Delta_1, Y_1, \cdots, X_K, A_K, \Delta_K, Y_K,\tilde{Y}\right\}$.

For simplicity, in the following simulation study, we use the linear regression model to fit $Q^k_s(L_k)$. It is worth noting that other complicated methods such as neural network can be adopted and analyzed in a similar manner, but it is not the focus of our work. As demonstrated in (\ref{Q-IPCW}) and (\ref{Q-IPCW2}), the $Q$-functions are estimated by the inverse probability censoring weighting (IPCW) least square method with different weights  $\frac{\Delta_k}{\hat{S}_C(\sum_{j=1}^kY_k)}$ and $\frac{\Delta}{\hat{S}_C(\tilde{Y})}$. Here, we use IPCW-I and IPCW-II to refer to the two methods, respectively. 
Intuitively, IPCW-I uses more information due to the additional information $Y_j, j=1,\cdots, k$ provided by the first data trajectory. Specifically, at stage $k$, IPCW-I uses the information from two groups of patients: (a) patients are not censored throughout all stages, (b) patients are censored after the $k$-th stage, while IPCW-II only uses the information of patients from the group (a) due to the lack of intermediate rewards. 


\subsection{Computation and Algorithm}

The classification problem (\ref{classification}) is an example dependent cost-sensitive multi-class weighted classification problem \citep{Elkan:2001}. Let $\{\tilde{L}_k, \hat{C}_k\}$ denote the classification problem, where $\tilde{L}_k=(L_{k1}, \cdots, L_{kn})^\top$ is the accrued history information up to stage $k$, and 
\[
\hat{C}_k=
\begin{bmatrix}
\hat{C}^k_1(L_{k1}) & \hat{C}^k_2(L_{k1}) & \cdots & \hat{C}^k_{m_k}(L_{k1}) \\
\hat{C}^k_1(L_{k2}) & \hat{C}^k_2(L_{k2}) & \cdots & \hat{C}^k_{m_k}(L_{k2}) \\ 
\vdots & \vdots &\ddots & \vdots\\
\hat{C}^k_1(L_{kn}) & \hat{C}^k_2(L_{kn}) & \cdots & \hat{C}^k_{m_k}(L_{kn}) \\
\end{bmatrix}
\]
is the cost matrix at the stage $k$. According to the definition of $\hat{C}_s^k(L_k)$, the cost corresponding to $s=\hat{l}_{m_k}^k(L_k)$ is zero. 
We can use the data space expansion \citep{Abe:2004} method to solve the classification problem $\{\tilde{L}_k, \hat{C}_k\}$.
Specifically, the problem $\{\tilde{L}_k, \hat{C}_k\}$ can be expressed individually as $\{(L_{ki}, \hat{C}_{ki})\}_{i=1}^n$, where $\hat{C}_{ki}=(\hat{C}_{1}^k(L_{ki}), \hat{C}_{2}^k(L_{ki}), \cdots, \hat{C}_{m_k}^k(L_{ki}))$ is the cost vector of individual $i$. We first define 
\begin{eqnarray*}
\hat{U}_{i,l}^k=\max_{1\leq s \leq m_k}\hat{C}_{s}^k(L_{ki})-\hat{C}_{l}^k(L_{ki}),~l=1,\cdots, m_k.
\end{eqnarray*} 
Then we modify the example dependent weighted classification problem $\{(L_{ki}, \hat{C}_{ki})\}_{i=1}^n$ into a regular weighted classification problem $\mathop{\cup}\limits_{i=1}^n\{L_{ki}, l, \hat{U}_{i,l}^k\}_{l=1}^{m_k}$, where $l$ is the label and $\hat{U}_{i,l}^k$ is the misclassification error.  The sample size of the data set after the expansion is $m_k$ times of the original one.
Any existing classifier can be used to solve the modified classification problem. As proved in \cite{Abe:2004}, solving the classification problem after data space expansion is equivalent to solving the unmodified one. In this paper, we adopt the CART, which possesses great advantages in interpretability and variable importance ranking, as the classifier. Hence the estimated optimal regime at each stage is a binary tree-type regime. 

We summarize our algorithm in Algorithm \ref{algorithm}, and denote it as CC-learning.
\begin{algorithm}[htb]
\caption{CC-learning algorithm}
\label{algorithm}
\begin{algorithmic}[1]
\REQUIRE ~~\\
$1.$ data $\{(X_{1i}, A_{1i}, \Delta_{1i}, \tilde{Y}_{1i}, \cdots, X_{\bar{K}i}, A_{\bar{K}i}, \Delta_{\bar{K}i}, Y_{\bar{K}i},\Delta_i, \tilde{Y}_i )\}_{i=1}^n$, or; \\
$1'.$ data $\{\left (X_{1i}, A_{1i}, \Delta_{1i},  \cdots, X_{\bar{K}i}, A_{\bar{K}i}, \Delta_{\bar{K}i},\Delta_i,  \tilde{Y}_i \right)\}_{i=1}^n$;\\
\ENSURE ~~\\
I. Construct the auxiliary problem and estimate the survival function of censoring time.\\
II. Backward induction based on the auxiliary problem. Let $k=K$.
\STATE If $k=K$, then $\hat{V}_{k+1}=\tilde{Y}$, otherwise estimate $\hat{V}_{k+1}(L_k)$ according to (\ref{M}) or (\ref{R});
\STATE Estimate the $Q$-function according to (\ref{Q-IPCW}) or (\ref{Q-IPCW2});
\STATE Estimate the contrast function according to (\ref{C-AIPW1}) or (\ref{C-AIPW2});
\STATE Conduct the data space expansion and construct the classification problem $\mathop{\cup}\limits_{i=1}^n\{L_{ki}, l, \hat{U}_{i,l}^k\}_{l=1}^{m_k}$;
\STATE Perform CRAT to estimate the optimal regime $\hat{g}_k^{opt}$ at stage $k$;
\STATE Stop algorithm when $k=1$; let $k=k-1$ when $k>1$, repeat previous steps.\\
III. Integrate the estimated regime at each stage to obtain the optimal dynamic treatment regime $\hat{g}^{opt}=(\hat{g}_1^{opt}, \cdots, \hat{g}^{opt}_K)$.
\end{algorithmic}
\end{algorithm}

\section{Simulation}\label{Simulation}
In this section, we conduct numerical studies to assess the performance of the proposed CC-learning algorithm. We consider two scenarios with the true optimal dynamic treatment regime of different types. In Scenario 1, the true optimal dynamic treatment regime belongs to the class of binary tree. In Scenario 2, the true optimal dynamic treatment regime belongs to a linear class. For both scenarios, there are two stages and each stage has three treatment options. 


\noindent{\bf Scenario 1.}
We first generate four covariates $X_1, X_2, X_3$ and $X_4$ independently from the normal distribution $N(0,1)$. Treatment $A_1$ is randomly generated according to the distribution $P(A_1=0)=\pi_{10}/\pi_{1s}, P(A_1=1)=\pi_{11}/\pi_{1s}, P(A_1=2)=\pi_{12}/\pi_{1s}$, where 
$\pi_{10}=1$, $\pi_{11}=\exp\{0.5-0.5X_3\}$, $\pi_{12}=\exp\{0.5X_4\}$ and $\pi_{1s}=\sum_{i=0}^2\pi_{1i}$. The optimal regime at the first stage is 
\begin{eqnarray*}
g_1^{opt}=\{X_1>-1\}\{(X_2>-0.5)+(X_2>0.5)\}.
\end{eqnarray*}
The first stage survival time $T_1$ is generated from 
\begin{eqnarray*}
T_1=\exp\{1.5-|1.5X_1+2|(A_1-g_1^{opt}(L_1))^2 +\epsilon_1\},
\end{eqnarray*}
where $\epsilon_1 \sim N(0,0.3^2)$.
Treatment $A_2$ is randomly generated according to the distribution $P(A_2=0)=\pi_{20}/\pi_{2s}, P(A_2=1)=\pi_{21}/\pi_{2s}, P(A_2=2)=\pi_{22}/\pi_{2s}$, where $\pi_{20}=1$, $\pi_{21}=\exp\{0.2T_1-1\}$, $\pi_{22}=\exp\{0.5X_4\}$ and $\pi_{2s}=\sum_{i=0}^2\pi_{2i}$. The optimal regime at the second stage is
\begin{eqnarray*}
g_2^{opt}=\{X_3>-1\}\{(T_1>0)+(T_1>2)\}.
\end{eqnarray*}
The second stage survival time $T_2$ is generated from
\begin{eqnarray*}
T_2=\exp\{1.26-|1.5X_3-2|(A_2-g_2^{opt}(L_2))^2 +\epsilon_2\},
\end{eqnarray*}
where $\epsilon_2 \sim N(0,0.3^2)$.
Then the total survival time is $T=T_1+T_2$. We generate the censoring time $C$ from the uniform distribution $U(0,C_0)$ and we can control the censoring rate by adjusting $C_0$. The censoring indicators $\Delta=I(T<C)$ and $\Delta_1=I(T_1<C)$ are generated accordingly.
Define $R$ as the indicator of whether a patient experiences a failure event at the first stage, with $R=0$ indicating that an event occurs and $R=1$ otherwise. We generate the index $R$ from the Bernoulli distribution with $P(R=1)=r$.  Then we generate the indicator for entering the second stage $\eta=R\Delta_1$. Finally, we generate survival time as follows: if $\eta=0$, then $Y_2=0$; if $\Delta_1=0$, then $Y_1=C$; if $\Delta=0$ and $\Delta_1=1$, then $Y_2=C-Y_1$; otherwise $Y_1=T_1$ and $Y_2=T_2$.

\noindent{\bf Scenario 2.}
We independently generate six covariates $X_1, \cdots, X_6$ from the normal distribution $N(0,1)$. Treatments $A_1$ and $A_2$ are generated in the same way as Scenario 1. Set the instrumental variables $A_{11}=I(A_1=1)$ and $A_{12}=I(A_1=2)$. And the optimal treatment regime of the first stage is set as follows: if $X_1-X_2<0$ and $X_1-X_2+X_3<0$, $g_1^{opt}=0$; if $X_1-X_2>0$ and $X_3<0$, $g_1^{opt}=1$; if $X_3>0$ and $X_1-X_2+X_3>0$, $g_1^{opt}=2$.
The first stage survival time $T_1$ is generated from
\begin{eqnarray*}
T_1=\exp\{1.5+0.5A_{11}(X_1-X_2)+0.5A_{12}(X_1-X_2+X_3)+\epsilon_1\},
\end{eqnarray*}
where $\epsilon_1 \sim N(0,0.3^2)$.
Set the instrumental variables $A_{21}=I(A_2=1)$ and $A_{22}=I(A_1=2)$.
And the optimal treatment regime of the second stage is set as follows: if $X_4<0$ and $X_4+X_5-X_6<0$, $g_2^{opt}=0$; if $X_4>0$ and $X_5-X_6<0$, $g_2^{opt}=1$; if $X_5-X_6>0$ and $X_4+X_5-X_6>0$, $g_2^{opt}=2$.
The second stage survival time $T_2$ is generated from
\begin{eqnarray*}
T_2=\exp\{1.5+0.5A_{21}X_4+0.5A_{22}(X_4+X_5-X_6)+\epsilon_2\},
\end{eqnarray*}
where $\epsilon_2 \sim N(0,0.3^2)$. The rest of the data generation is the same as Scenario 1.

 In this simulation, we will investigate the impacts of different sample sizes, censoring rates, IPCW methods (IPCW-I and IPCW-II) and backward induction methods (R-method and D-method) on the performance of our proposed CC-learning algorithm. In Scenario 1, $C_0$ is set as 35 and 18 to achieve the censoring rates of 10\% and 20\%. In Scenario 2, $C_0=115$ and 55 so that the corresponding censoring rates are 10\% and 20\%, respectively. In Scenario 1, the true models for the outcome are neither linear nor log-linear models. Hence the $Q$-function is always mis-specified (QF). In order to achieve consistency, we set the propensity score models to be always correctly specified (PST). In Scenario 2, we use accelerated failure time (AFT) model for the outcome regression, thus the $Q$-functions are correctly specified (QT). We also specify the correct propensity score model (PST) for simplicity. In both scenarios, we set $r$ to 1 and 0.85, and consider different sample sizes $n=300$, 500 and 1000. The simulation results based on 100 Monte Carlo replicates are presented in Tables \ref{simulation-tree} and \ref{simulation-nontree}.
 
\begin{table}[]
\caption{Simulation results of Scenario 1}
\normalsize
\centering
\label{simulation-tree}
\begin{tabular}{lllllllllll}
\hline
\multicolumn{1}{c}{\multirow{2}{*}{$n$}} & \multicolumn{1}{c}{AA1} & \multicolumn{1}{c}{AA2} & \multicolumn{1}{c}{AA} & \multicolumn{1}{c}{$\hat{V}$} & \multicolumn{1}{c}{$V^{opt}$} & \multicolumn{1}{c}{AA1} & \multicolumn{1}{c}{AA2} & \multicolumn{1}{c}{AA} & \multicolumn{1}{c}{$\hat{V}$} & \multicolumn{1}{c}{$V^{opt}$} \\
\multicolumn{1}{c}{}                     & \multicolumn{5}{c}{CR=10\%}                                                                                                             & \multicolumn{5}{c}{CR=20\%}                                                                                                             \\ \hline
\multicolumn{11}{c}{IPCW-I~~ PST~~ QF~~ D-method~~ $r=1$}                                                                                                                                                                                                                                                                             \\
300                                      & 0.907                   & 0.906                   & 0.880                  & 7.406                       & 8.007                        & 0.843                   & 0.826                   & 0.782                  & 6.953                      & 8.007                        \\
500                                      & 0.935                   & 0.940                   & 0.921                  & 7.606                       & 8.007                        & 0.895                   & 0.900                   & 0.869                  & 7.348                       & 8.007                        \\
1000                                     & 0.970                   & 0.976                   & 0.968                  & 7.836                      & 8.007                        & 0.947                   & 0.958                   & 0.943                  & 7.700                       & 8.007                        \\
\multicolumn{11}{c}{IPCW-II~~ PST~~ QF~~ D-method~~ $r=1$}                                                                                                                                                                                                                                                                             \\
300                                      & 0.882                   & 0.888                   & 0.856                  & 7.266                       & 8.007                        & 0.786                   & 0.792                   & 0.733                  & 6.659                     & 8.007                        \\
500                                      & 0.912                   & 0.924                   & 0.899                  & 7.478                  & 8.007                        & 0.859                   & 0.875                   & 0.835                  & 7.158                       & 8.007                        \\
1000                                     & 0.961                   & 0.969                   & 0.959                  & 7.784                       & 8.007                        & 0.924                   & 0.940                   & 0.920                  & 7.566                       & 8.007                        \\
\multicolumn{11}{c}{IPCW-I~~ PST~~ QF~~ R-method~~ $r=1$}                                                                                                                                                                                                                                                                             \\
300                                      & 0.889                   & 0.895                   & 0.863                  & 7.307                       & 8.007                        & 0.837                   & 0.825                   & 0.779                  & 6.778                       & 8.007                        \\
500                                      & 0.927                   & 0.935                   & 0.914                  & 7.564                       & 8.007                        & 0.891                   & 0.898                   & 0.865                  & 7.330                       & 8.007                        \\
1000                                     & 0.964                   & 0.972                   & 0.961                  & 7.800                       & 8.007                        & 0.940                   & 0.953                   & 0.936                  & 7.664                       & 8.007                        \\
\multicolumn{11}{c}{IPCW-II~~ PST~~ QF~~ R-method~~ $r=1$}                                                                                                                                                                                                                                                                             \\
300                                      & 0.859                   & 0.875                   & 0.835                  & 7.143                       & 8.007                        & 0.771                   & 0.784                   & 0.719                  & 6.572                       & 8.007                        \\
500                                      & 0.900                   & 0.916                   & 0.887                  & 7.415                       & 8.007                        & 0.843                   & 0.864                   & 0.819                  & 7.064                       & 8.007                        \\
1000                                     & 0.953                   & 0.963                   & 0.950                  & 7.736                       & 8.007                        & 0.912                   & 0.932                   & 0.908                  & 7.502                       & 8.007                        \\
\multicolumn{11}{c}{IPCW-I~~ PST~~ QF~~ D-method~~ $r=0.85$}                                                                                                                                                                                                                                                                           \\
300                                      & 0.899                   & 0.893                   & 0.865                  & 7.341                       & 8.007                        & 0.837                   & 0.796                   & 0.750                  & 7.016                       & 8.007                        \\
500                                      & 0.934                   & 0.941                   & 0.922                  & 7.607                       & 8.007                        & 0.902                   & 0.889                   & 0.861                  & 7.346                      & 8.007                        \\
1000                                     & 0.973                   & 0.977                   & 0.969                  & 7.764                       & 8.007                        & 0.946                   & 0.949                   & 0.934                  & 7.670                       & 8.007                        \\
\multicolumn{11}{c}{IPCW-II~~ PST~~ QF~~ D-method~~ $r=0.85$}                                                                                                                                                                                                                                                                           \\
300                                      & 0.866                   & 0.873                   & 0.835                  & 7.169                      & 8.007                        & 0.778                   & 0.761                   & 0.700                  & 6.534                       & 8.007                        \\
500                                      & 0.919                   & 0.930                   & 0.906                  & 7.520                      & 8.007                        & 0.871                   & 0.867                   & 0.832                  & 7.170                       & 8.007                        \\
1000                                     & 0.963                   & 0.969                   & 0.959                  & 7.786                       & 8.007                        & 0.924                   & 0.933                   & 0.913                  & 7.546                       & 8.007                        \\
\multicolumn{11}{c}{IPCW-I~~ PST~~ QF~~ R-method~~ $r=0.85$}                                                                                                                                                                                                                                                                           \\
300                                      & 0.883                   & 0.884                   & 0.850                  & 7.256                      & 8.007                        & 0.822                   & 0.788                   & 0.738                  & 6.765                      & 8.007                        \\
500                                      & 0.929                   & 0.937                   & 0.916                  & 7.578                       & 8.007                        & 0.899                   & 0.887                   & 0.858                  & 7.326                       & 8.007                        \\
1000                                     & 0.968                   & 0.973                   & 0.965                  & 7.816                      & 8.007                        & 0.945                   & 0.948                   & 0.933                  & 7.663                       & 8.007                        \\
\multicolumn{11}{c}{IPCW-II~~ PST~~ QF~~ R-method~~ $r=0.85$}                                                                                                                                                                                                                                                                           \\
300                                      & 0.854                   & 0.865                   & 0.824                  & 7.100                       & 8.007                        & 0.771                   & 0.758                   & 0.696                  & 6.650                       & 8.007                        \\
500                                      & 0.907                   & 0.922                   & 0.895                  & 7.456                       & 8.007                        & 0.858                   & 0.859                   & 0.819                  & 7.102                       & 8.007                        \\
1000                                     & 0.953                   & 0.963                   & 0.950                  & 7.733                       & 8.007                        & 0.915                   & 0.927                   & 0.903                  & 7.494                      & 8.007                        \\ \hline
\end{tabular}
\end{table}

\begin{table}[]
\caption{Simulation results of Scenario 2}
\centering
\normalsize 
\label{simulation-nontree}
\begin{tabular}{lllllllllll}
\hline
\multirow{2}{*}{$n$} & AA1   & AA2   & AA    & $\hat{V}$ & $V^{opt}$ & AA1   & AA2   & AA    & $\hat{V}$ & $V^{opt}$ \\
                     & \multicolumn{5}{c}{CR=10\%}                   & \multicolumn{5}{c}{CR=20\%}                   \\ \hline
\multicolumn{11}{c}{IPCW-I~~ PST~~ QT~~ D-method~~ $r=1$}                                                                     \\
300                  & 0.578 & 0.723 & 0.417 & 12.775    & 14.302    & 0.527 & 0.682 & 0.360 & 12.291    & 14.302    \\
500                  & 0.587 & 0.750 & 0.440 & 12.912    & 14.302    & 0.546 & 0.690 & 0.376 & 12.400    & 14.302    \\
1000                 & 0.621 & 0.760 & 0.473 & 13.102    & 14.302    & 0.576 & 0.725 & 0.417 & 12.709    & 14.302    \\
\multicolumn{11}{c}{IPCW-II~~ PST~~ QT~~ D-method~~ $r=1$}                                                                     \\
300                  & 0.539 & 0.733 & 0.394 & 12.629    & 14.302    & 0.480 & 0.682 & 0.327 & 12.001    & 14.302    \\
500                  & 0.559 & 0.744 & 0.415 & 12.743    & 14.302    & 0.494 & 0.691 & 0.340 & 12.160    & 14.302    \\
1000                 & 0.582 & 0.761 & 0.444 & 12.928    & 14.302    & 0.523 & 0.725 & 0.378 & 12.506    & 14.302    \\
\multicolumn{11}{c}{IPCW-I~~ PST~~ QT~~ R-method~~ $r=1$}                                                                     \\
300                  & 0.592 & 0.733 & 0.433 & 12.840    & 14.302    & 0.545 & 0.682 & 0.371 & 12.338    & 14.302    \\
500                  & 0.588 & 0.744 & 0.437 & 12.893    & 14.302    & 0.553 & 0.691 & 0.380 & 12.436    & 14.302    \\
1000                 & 0.628 & 0.761 & 0.478 & 13.142    & 14.302    & 0.591 & 0.725 & 0.428 & 12.793    & 14.302    \\
\multicolumn{11}{c}{IPCW-I~~ PST~~ QT~~ R-method~~ $r=1$}                                                                     \\
300                  & 0.540 & 0.733 & 0.394 & 12.637    & 14.302    & 0.479 & 0.683 & 0.327 & 12.014    & 14.302    \\
500                  & 0.562 & 0.744 & 0.417 & 12.767    & 14.302    & 0.501 & 0.691 & 0.344 & 12.183    & 14.302    \\
1000                 & 0.593 & 0.761 & 0.452 & 12.980    & 14.302    & 0.528 & 0.725 & 0.382 & 12.539    & 14.302    \\
\multicolumn{11}{c}{IPCW-I~~ PST~~ QT~~ D-method~~ $r=0.85$}                                                                   \\
300                   & 0.576 & 0.732 & 0.422 & 12.757    & 14.302  & 0.526 & 0.663 & 0.349 & 12.192    & 14.302   \\
500                   & 0.587 & 0.748 & 0.439 & 12.846    & 14.302   & 0.547 & 0.681 & 0.372 & 12.355   & 14.302    \\
1000                  & 0.618 & 0.756 & 0.466 & 13.110    & 14.302   & 0.585 & 0.714 & 0.416 & 12.713    & 14.302     \\
\multicolumn{11}{c}{IPCW-II~~ PST~~ QT~~ D-method~~ $r=0.85$}                                                                   \\
300                  & 0.532 & 0.733 & 0.390 & 12.583    & 14.302    & 0.483 & 0.664 & 0.321 & 12.004    & 14.302    \\
500                  & 0.548 & 0.748 & 0.410 & 12.695    & 14.302    & 0.499 & 0.682 & 0.339 & 12.157    & 14.302    \\
1000                 & 0.593 & 0.756 & 0.448 & 12.973    & 14.302    & 0.527 & 0.714 & 0.375 & 12.475    & 14.302    \\
\multicolumn{11}{c}{IPCW-I~~ PST~~ QT~~ R-method~~ $r=0.85$}                                                                   \\
300                  & 0.577 & 0.732 & 0.422 & 12.773    & 14.302    & 0.536 & 0.664 & 0.355 & 12.206    & 14.302    \\
500                  & 0.596 & 0.748 & 0.445 & 12.888    & 14.302    & 0.552 & 0.681 & 0.375 & 12.393    & 14.302    \\
1000                 & 0.627 & 0.756 & 0.473 & 13.138    & 14.302    & 0.589 & 0.714 & 0.419 & 12.723    & 14.302    \\
\multicolumn{11}{c}{IPCW-I~~ AIPW~~ PST~~ QT~~ R-method~~ $r=0.85$}                                                                   \\
300                  & 0.532 & 0.733 & 0.390 & 12.591    & 14.302    & 0.490 & 0.664 & 0.325 & 12.051    & 14.302    \\
500                  & 0.550 & 0.748 & 0.411 & 12.698    & 14.302    & 0.508 & 0.682 & 0.346 & 12.204    & 14.302    \\
1000                 & 0.597 & 0.756 & 0.450 & 12.979    & 14.302    & 0.528 & 0.714 & 0.375 & 12.460    & 14.302    \\ \hline
\end{tabular}
\end{table}

When evaluating the proposed algorithm, of the most interest is $\hat{V}$, which is the  mean survival time following the estimated optimal dynamic treatment regime. $V^{opt}$ is the mean survival time following the true optimal dynamic treatment regime. If $\hat{g}^{opt}$ is satisfactory, $\hat{V}$ should be close to $V^{opt}$. Besides, we show the assignment accuracy, say, the proportion of treatments assigned by the estimated optimal dynamic treatment regime consistent with the true optimal treatments. AA1, AA2 and AA are the first stage, the second stage and overall assignment accuracy, respectively. A higher assignment accuracy means better estimation of optimal dynamic treatment regimes. For assessing the generalization ability of the estimated optimal dynamic treatment regime, $\hat{V}$, AA1, AA2 and AA are computed on the testing set of sample size 10000. We also compute the mean survival time following random assignment $\tilde{V}$ on the testing set to better demonstrate the effect of the estimated dynamic treatment regime. In scenario 1, $\tilde{V}=3.518$; and in Scenario 2, $\tilde{V}=9.870$.

For Scenario 1, the $\hat{V}$ and assignment accuracy generally increase with the sample size. The increase of the censoring rate has negative impacts on the performance of CC-learning method. From the perspective of $\hat{V}$, AA1, AA2 and AA, IPCW-I outperforms IPCW-II, which is consistent with the analysis in section \ref{Clearning}. With $r$ decreasing from 1 to 0.85, the mean survival time and the assignment accuracy decrease to some extent due to the fact that fewer people enter the second stage. In most cases, the D-method is slightly better than the R-method. Overall, the performance of CC-learning method in Scenario 1 is satisfactory with $\hat{V}$ much larger than $\tilde{V}=3.518$ and assignment accuracy close to 1.

For Scenario 2, $\hat{V}$ is much larger than $\tilde{V}=9.870$ for all cases. Larger sample size, smaller censoring rate or smaller $r$ result in better performances. The IPCW-I is better than IPCW-II, but the D-method is worse than the R-method in Scenario 2.
It is worth noting that, though the assignment accuracy is much lower than that of Scenario 1, the $\hat{V}$ is still close to $V^{opt}$. Actually, when the optimal treatment strategy is non-tree-like, CART fails to approximate the true optimal decision, but the effects of treatment determined by the estimated treatment strategy is still good.
In this scenario, the goal of the optimal dynamic treatment regime is to maximize the mean survival time rather than the assignment accuracy. The maximization of the outcome of interest on average can be achieved at the sacrifice of the assignment accuracy of those patients with relatively small treatment effects.

\section{Real data analysis}\label{Data analysis}

We apply the proposed method to the advanced colorectal cancer study data described in Section \ref{data}. The details of the covariates are shown in Table \ref{covariates}.

\begin{table}[]
\caption{The details of the covariates}
    \centering
    \begin{tabular}{c|l}
    \hline
    Variable & Description\\
    \hline
       ${\rm Time}_1$ & Treatment time at the first stage\\
        Gender  & 0: female, 1: male \\
        Age & 0: $<75$, 1: $\geq$75\\
        Site & Site of pathological changes, 0: left, 1: right\\
        Metastasis & 1: high prognostic impact metastasis, 2: moderately prognostic impact metastasis\\
        Genotyping &  1: KRAS, 2: RAS, 3: BRAF, 4: unspecified\\
        ECOG1 & ECOG score at the first stage, 0: $<2$, 1: $\geq 2$    \\
        ECOG2 & ECOG score at the SECOND stage, 0: $<2$, 1: $\geq 2$\\
        Stage &Medication stage of initial diagnosis,\\
        &1: primary treatment and first-line, 2: second-line, 3: third-line and above\\
    \hline
    \end{tabular}
    \label{covariates}
\end{table}

\begin{figure}[htbp]
\centering
\includegraphics[width=15cm]{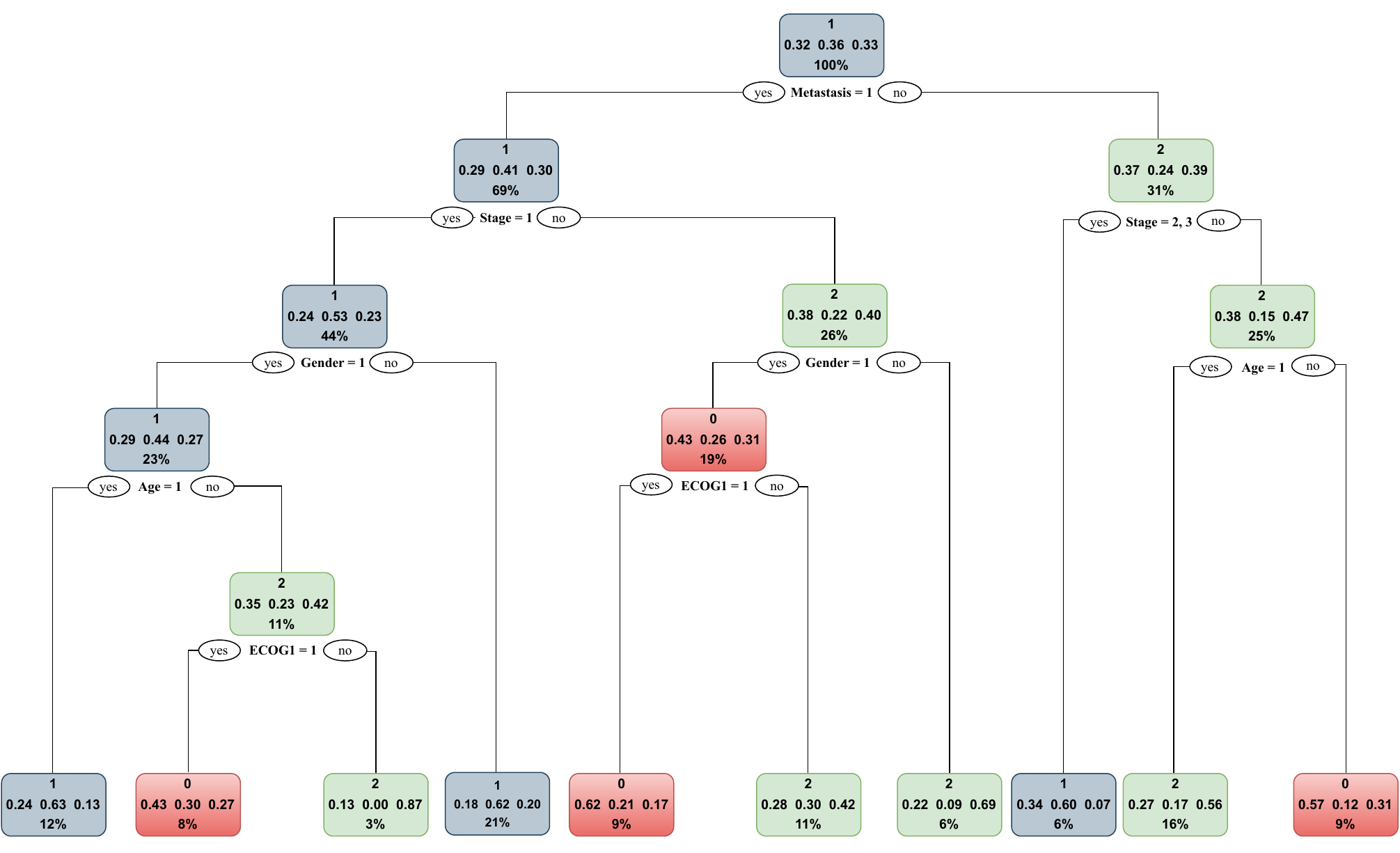}
\caption{The estimated optimal regime at the first stage.}
\label{dtr1}
\end{figure}

\begin{figure}[htbp]
\centering
\includegraphics[width=12cm]{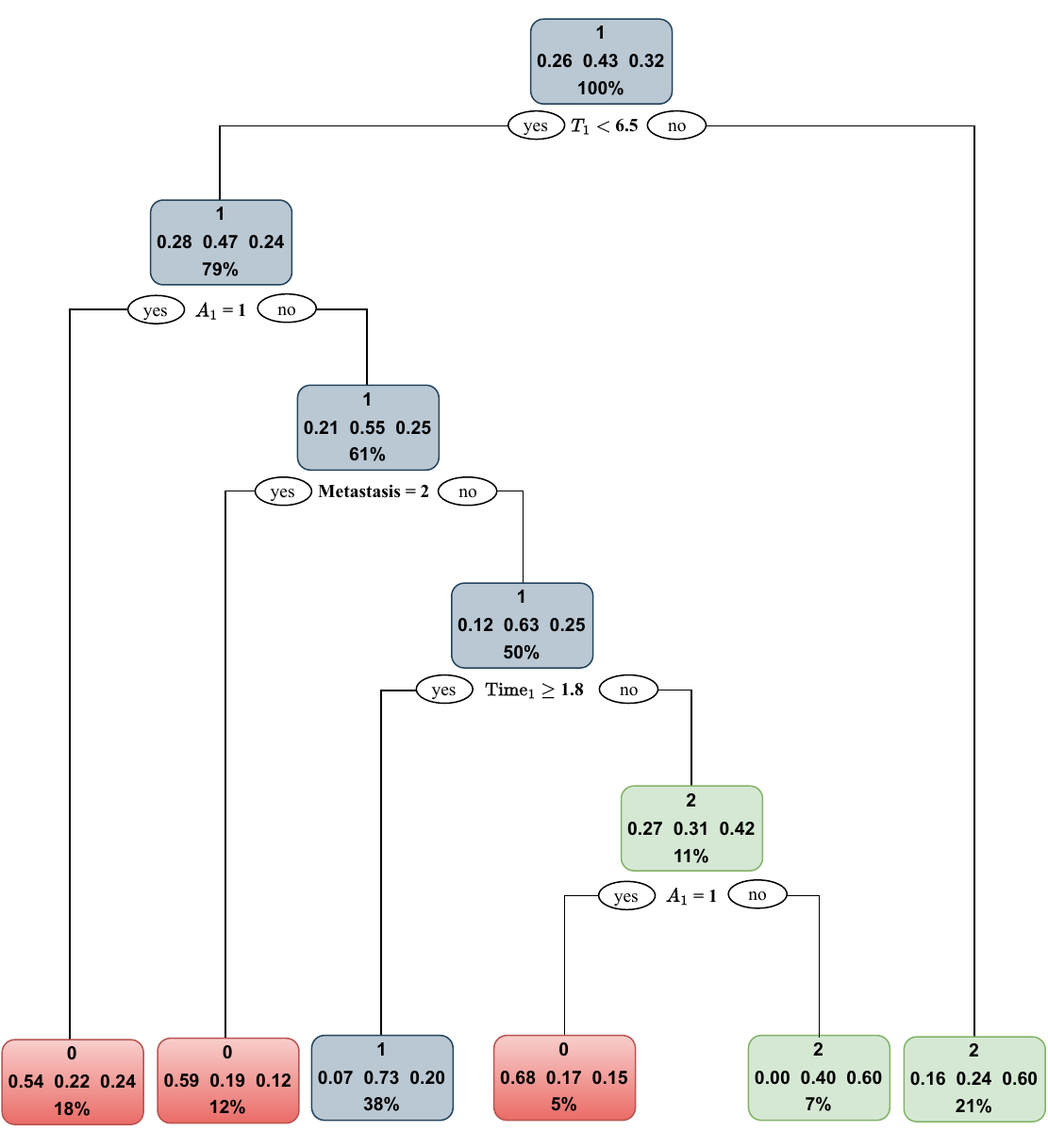}
\caption{The estimated optimal regime at the second stage.}
\label{dtr2}
\end{figure}

We use the AFT model to fit the $Q$-function, and softmax regression to fit the propensity score model. Specifically, denote the main effect term in the AFT model as ``tf.mod'', the interaction term in the AFT model as ``blip.mod'', propensity score model as ``treat.mod'' and classification model as ``classification.mod''. The specific model setting is presented in table \ref{model-setting}. We use IPCW-II to deal with censoring and D-method for backward induction. After applying our proposed CC-learning algorithm, we obtain the following decision tree for the optimal dynamic treatment regime of advanced colorectal cancer with traditional Chinese medicine. The direction of each branch of the tree is determined by decision points, and multiple decision points collectively determine the final action for the strategy. The decision tree diagram ultimately outputs the endpoint of the decision, which is the treatment plan influenced by the factors involved in the decision. The estimated optimal regimes at the first and second stages are presented in Figure \ref{dtr1} and \ref{dtr2}, respectively.

Tree-based decisions offer intuitive interpretability. In each box of Figure \ref{dtr1} and Figure \ref{dtr2}, the first row represents the treatment allocation result. The second row of vectors represents the probabilities of individuals with the characteristics of this box being assigned to each treatment, with the highest probability corresponding to the final assigned treatment. The percentage in the third row represents the proportion of individuals in the training set falling into the treatment we recommend out of the total number of individuals. Starting at the root in Figure \ref{dtr1}, conditions like Metastasis=1 guide the path through left or right nodes, assigning treatment choices. This tree-like approach resembles clinical decision-making, simplifying understanding. Taking the first-stage treatment strategy as an example, for patients with moderately prognostic impact metastasis (Metastasis=2) and second-line or above medication stage of initial diagnosis (Stage=2,3), a combination of Chinese and Western medicine is recommended. For patients with high prognostic impact metastasis (Metastasis=1) and first-line medication stage of initial diagnosis (Stage=1), the optimal first-stage treatment is also the combination of Chinese and Western medicine. For male patients with highly prognostic impact metastasis (Metastasis=1), no prior tumor-related treatments (Stage=1), the optimal treatment is Western medicine assisted with traditional Chinese medicine. 

 \begin{table}[]
\centering
\label{model}
\footnotesize
\caption{Models}
\label{model-setting}

\begin{tabular}{cc}
\hline
\multicolumn{1}{l}{Model}                 & Setting                                                                              \\ \hline
\multicolumn{1}{l}{tf.mod}             &                                                                                 \\
Stage 1                                 & Age + Gender + Site + Metastasis + Genotyping + Stage + ECOG1                   \\
Stage 2                                   & Age + Gender + Site + Metastasis + Genotyping + Stage + ECOG1 + ${\rm Time}_1$ + $A_1$ + ECOG2 \\
\multicolumn{1}{l}{blip.mod}           &                                                                                 \\
Stage 1                                      & Age + Gender + Site + Metastasis + Genotyping + Stage + ECOG1                   \\
Stage 2                                & Age + Gender + Site + Metastasis + Genotyping + Stage + ECOG1 + $A_1$ + ECOG2      \\
\multicolumn{1}{l}{treat.mod}          &                                                                                 \\
Stage 1                                       & Age + Gender + Site + Metastasis + Genotyping + Stage + ECOG1                   \\
Stage 2                                 & Age + Gender + Site + Metastasis + Genotyping + Stage + ECOG1 + ${\rm Time}_1$ + $A_1$ + ECOG2 \\
\multicolumn{1}{l}{classification.mod} &                                                                                 \\
Stage 1                                       & Age + Gender + Site + Metastasis + Genotyping + Stage + ECOG1                   \\
Stage 2                                 & Age + Gender + Site + Metastasis + Genotyping + Stage + ECOG1 + ${\rm Time}_1$ + $A_1$ + ECOG2\\ \hline
\end{tabular}
\end{table}

\section{Conclusions}\label{Conclusions}
In this paper, we extend the C-learning framework to multiple treatments and survival data cases. Our proposed CC-learning algorithm incorporates the data space expansion method to solve the example dependent cost-sensitive classification problem and can be used to estimate the optimal dynamic treatment regime maximizing mean survival time. Simulation results show that the CC-learning algorithm has desired performance no matter whether the true optimal dynamic treatment regime is in the pre-specified class $\mathcal{G}$.
The dynamic treatment regime estimated by CC-learning method is remarkably parsimonious and interpretable, which allows researchers to create novel diagnostics or guide therapeutic advances for patients. It is worth noting that other classifiers besides CART can be used to construct the optimal dynamic treatment regime based on the practical demand.

The performance of CC-learning method depends heavily on the specification of $Q$-function. A misspecified $Q$-function will not only affect the estimation of contrast function but also the estimation of the value function, hence the backward induction. In that case, even if the propensity score model is correctly specified, the AIPW estimator would not be consistent. Actually, it is a common problem for methods based on backward induction. Thus, one possible improvement is to use nonparametric methods for outcome regression when the sample size is large enough. As demonstrated in the Scenario 2 of the simulation study, when the true optimal dynamic treatment regime is not in the pre-specified class $\mathcal{G}$, the assignment accuracy can be very low, which means a large proportion of patients cannot be assigned the best treatment. Considering from the perspective of fairness, RCSCLSurv is not that satisfactory. Thus, another possible improvement and extension is that we constrain the assignment accuracy to be larger than a threshold while maximizing the mean survival time to consider fairness, which is an interest problem that warrants future study.

\section*{Acknowledgements}
Lin's work was supported by the MOE Project of Key Research Institute of Humanities and
Social Sciences (22JJD910001). Yang's work was support by the Capital Health Development Scientific Research Project (2022-1-4171).



\begin{appendix}

\section*{}

\subsection{Proof of Theorem \ref{th1}}
We first start at stage $K$. On the one hand, we have
\begin{align}
V_K(L_K)&=E(\tilde{Y}^*(L_K, g_K^{opt})|L_K)\notag \\
&=E(\tilde{Y}|L_K,A_K=g_K^{opt})\label{pro1-1} \\ 
&=Q_K(L_K,A_K=1)I(g_K^{opt}=1) + \cdots + \notag \\ 
&\quad Q_K(L_K,A_K=m_K)I(g_K^{opt}=m_K). \notag
\end{align}
The equation is guaranteed by the consistency assumption and NUCA. On the other hand, we have
\begin{align}
&E(\tilde{Y}+(Q_K(L_K,g_K^{opt}) - Q_K(L_K,A_K))|L_K)\notag \\
&=E(Q_K(L_K,A_K) + (Q_K(L_K,g_K^{opt}) - Q_K(L_K,A_K))|L_K)\notag \\
&=Q_K(L_K,A_K=1)I(g_K^{opt}=1)+\cdots +Q_K(L_K,)I(g_K^{opt}=m_K) \notag \\ 
&= V_K(L_K). \notag
\end{align}
Similarly, at the stage $k,K-1\leq k \leq 1$, on the one hand we have
\begin{align}
V_k(L_k)&=E(\tilde{Y}^*(L_k, g_k^{opt}, g_{k+1}^{opt}, \cdots, g_K^{opt})|L_k)\notag \\
&=E(\tilde{Y}(L_k, g_k^{opt}, g_{k+1}^{opt}, \cdots, g_K^{opt})|L_k, A_k=g_k^{opt})\notag \\
&=E(V_{k+1}(L_k,g_{k}^{opt}(L_k),X_{k+1})|L_k,A_k=g_k^{opt})\notag \\ 
&=Q_k(L_k,A_k=1)I(g_k^{opt}=1) + \cdots + Q_k(L_k,A_k=m_k)I(g_k^{opt}=m_k). \notag
\end{align}
On the other hand, we have
\begin{align}
&E(V_{k+1}(L_{k+1})+(Q_k(L_k,g_k^{opt})-Q_k(L_k,A_k))|L_k)\notag \\
&=E(Q_k(L_k, A_k)+(Q_k(L_k,g_k^{opt})-Q_k(L_k,A_k))|L_k)\notag \\
&=Q_k(L_k,A_k=1)I(g_k^{opt}=1)+\cdots +Q_k(L_k,A_k=m_k)I(g_k^{opt}=m_k)\notag \\
&=V_k(L_k).\notag
\end{align}
Then the result is proved.

\subsection{Proof of Theorem \ref{th2}}
The backward induction and value function are expressed in terms of the observed data in Section \ref{Methodology}. Here, we redefine the backward induction and value function based on the potential outcome. For the sake of clarity, here we use abbreviations or no as appropriate.

At the stage $K$, define
\begin{eqnarray*}
g_K^{opt}(\bar{x}_K, \bar{a}_{K-1})=\argmax_{a_K\in \mathcal{A}_K}E\left \{\tilde{Y}^*(\bar{x}_K,\bar{a}_{K-1},a_K)|\bar{X}^*_K(\bar{a}_{K-1})=\bar{x}_K\right \},
\end{eqnarray*}
\begin{eqnarray*}
V_K(\bar{x}_K, \bar{a}_{K-1}) = \max_{a_K\in \mathcal{A}_K}E\left \{\tilde{Y}^*(\bar{x}_K, \bar{a}_{K-1},a_K)|\bar{X}^*_K(\bar{a}_{K-1})=\bar{x}_K\right \}.
\end{eqnarray*}
For $k=K-1,\cdots,2$ and $\forall$ $\bar{x}_k \in \bar{\mathcal{X}}_k$ and $\forall$ $\bar{a}_{k-1}\in \bar{\mathcal{A}}_{k-1}$, define
\begin{eqnarray*}
g_k^{opt}(\bar{x}_k, \bar{a}_{k-1})=\argmax_{a_k\in \mathcal{A}_k}E[V_{k+1}\left\{\bar{x}_k, X^*_{k+1}(\bar{a}_{k-1},a_k),\bar{a}_{k-1},a_k\right\}|\bar{X}^*_k(\bar{a}_{k-1})=\bar{x}_k],
\end{eqnarray*}
\begin{eqnarray*}
V_k(\bar{x}_k,\bar{a}_{k-1})=\max_{a_k\in \mathcal{A}_k}E[V_{k+1}\left\{\bar{x}_k, X^*_{k+1}(\bar{a}_{k-1},a_k),\bar{a}_{k-1},a_k\right\}|\bar{X}^*_k(\bar{a}_{k-1})=\bar{x}_k].
\end{eqnarray*}
For $k=1$, $x_1\in \mathcal{X}_1$, define
\begin{eqnarray*}
g_1^{opt}(x_1)=\argmax_{a_1 \in \mathcal{A}_1}E[V_2\left\{x_1,X^*_2(a_1),a_1\right\}|X_1=x_1],
\end{eqnarray*}
\begin{eqnarray*}
V_1(x_1)=\max_{a_1 \in \mathcal{A}_1}E[V_2\left\{x_1,X^*_2(a_1),a_1\right\}|X_1=x_1].
\end{eqnarray*}
Recall that $L_k \equiv (\bar{X}_k,\bar{A}_{k-1})$, we have
\begin{eqnarray*}
V_K(L_K)=E\left\{\tilde{Y}^*(\bar{A}_{K-1},g_{K}^{opt})|L_K\right\},
\end{eqnarray*}
\begin{eqnarray*}
V_k(L_k)=E\left\{\tilde{Y}^*(\bar{A}_{k-1},g_k^{opt}, g_{k+1}^{opt},\cdots,g_K^{opt}|L_k \right\},k=K-1,\cdots,1.
\end{eqnarray*}
We first construct the optimal regime at stage $K$, according to the definition, we have
\begin{align}
g_K^{opt}&=\argmin_{g_K\in \mathcal{G}_K}E(\tilde{Y}^*(L_K,g_K))\notag \\
&=\argmax_{g_K\in \mathcal{G}_K}E[E(\tilde{Y}^*(L_K,g_K)|L_K)] \notag \\
&=\argmax_{g_K\in \mathcal{G}_K}E[E(\tilde{Y}^*(L_K,g_K)|L_K, A_K=g_K)]\label{th1-1} \\
&=\argmax_{g_K\in \mathcal{G}_K}E[E(\tilde{Y}|L_K, A_k=g_K)] \label{th1-2} \\
&=\argmax_{g_K\in \mathcal{G}_K}E[E(\tilde{Y}|L_K, A_K=1)I(g_K=1) + \cdots + \notag \\
&\qquad \qquad \quad E(\tilde{Y}|L_K, A_K=m_K)I(g_K=m_K)] \notag  \\ 
&=\argmax_{g_K\in \mathcal{G}_K}E[Q_K(L_K,A_K=1)I(g_K=1) + \cdots + \notag \\
&\qquad \qquad \quad Q_K(L_K,A_K=m_K)I(g_K=m_K)] \notag \\
&=\argmin_{g_K\in \mathcal{G}_K}E[-Q_K(L_K,A_K=1)I(g_K=1) - \cdots - \notag \\
&\qquad \qquad \quad Q_K(L_K,A_K=m_K)I(g_K=m_K)] \notag \\ 
&=\argmin_{g_K\in \mathcal{G}_K}E[(Q_K(L_K,l_{m_K}^K(L_K))-Q_K(L_K,A_K=1))I(g_K=1) + \cdots + \notag\\
&\qquad \qquad \quad (Q_K(L_K,l_{m_K}^K(L_K))-Q_K(L_K,A_K=m_K))I(g_K=m_K)] \notag \\
&=\argmin_{g_K\in \mathcal{G}_k}E[C_{g_K}(L_K)I(g_K\neq l_{m_K}^K(L_K)].\notag
\end{align}
The equation (\ref{th1-1}) is guaranteed by the NUCA, and the equation (\ref{th1-2}) is guaranteed by the consistency assumption. Recursively, at the stage $k  = K-1,\cdots, 1$, we have
\begin{align}
g_k^{opt}&=\argmin_{g_k\in \mathcal{G}_k}E(\tilde{Y}^*(L_k,g_k, g_{k+1}^{opt}, \cdots, g_K^{opt}))\notag \\
&=\argmax_{g_k\in \mathcal{G}_k}E[E(\tilde{Y}^*(L_k,g_k, g_{k+1}^{opt}, \cdots, g_K^{opt})|L_K] \notag \\
&=\argmax_{g_k\in \mathcal{G}_k}E[E(\tilde{Y}^*(L_k,g_k), g_{k+1}^{opt}, \cdots, g_K^{opt})|L_k, A_k=g_k],\label{th1-3}
\end{align}
and 
\begin{align}
&E(\tilde{Y}^*(L_k,g_k, g_{k+1}^{opt}, \cdots, g_K^{opt})|L_k, A_k=g_k) \notag\\
&=E[E(\tilde{Y}^*(L_{k},g_k, g_{k+1}^{opt}, \cdots, g_K^{opt})|L_k,A_k=g_k,X_{k+1}^*(L_k,g_k)|L_k,A_k=g_k]\notag \\
&=E[E(\tilde{Y}^*(L_{k},g_k, g_{k+1}^{opt}, \cdots, g_K^{opt})|L_k,A_k=g_k,X_{k+1})|L_k,A_k=g_k]
\notag \\
&=E[V_{k+1}(L_k,g_k,X_{k+1})|L_k,A_k=g_k].\label{th1-4}
\end{align}
The equation (\ref{th1-3}) is guaranteed by the NUCA, and the equation (\ref{th1-4}) is guaranteed by the definition of value function. Substitute (\ref{th1-4}) into (\ref{th1-3}), we have
\begin{align}
g_k^{opt}&=\argmax_{g_k\in \mathcal{G}_k}E[E(V_{k+1}(L_k,g_k,X_{k+1})|L_k,A_k=g_k)] \notag \\
&=\argmax_{g_k\in \mathcal{G}_k}E[Q_k(L_k, A_k=g_k)]\label{th1-5} \\
&=\argmax_{g_k\in \mathcal{G}_k}E[Q_k(L_k,A_k=1)I(g_k=1) + \cdots + \notag \\
&\qquad \qquad \quad Q_k(L_k,A_k=m_k)I(g_k=m_k)] \notag \\
&=\argmin_{g_k\in \mathcal{G}_k}E[-Q_k(L_k,A_k=1)I(g_k=1) - \cdots - \notag \\
&\qquad \qquad \quad Q_K(L_k,A_k=m_k)I(g_k=m_k)] \notag \\ 
&=\argmin_{g_k\in \mathcal{G}_k}E[(Q_k(L_k,l_{m_k}^k(L_k))-Q_k(L_k,A_k=1))I(g_k=1) + \cdots + \notag\\
&\qquad \qquad \quad (Q_k(L_k,l_{m_k}^k(L_k))-Q_k(L_k,A_k=m_k))I(g_k=m_k)] \notag \\
&=\argmin_{g_k\in \mathcal{G}_k}E[C_{g_k}(L_k)I(g_k\neq l_{m_k}^k(L_k)].\notag
\end{align}
The equation (\ref{th1-5}) is guaranteed by the definition of $Q$-function. Then theorem \ref{th1} is proved.

\end{appendix}
\bibliographystyle{imsart-nameyear} 
\bibliography{ref}       

\end{document}